\documentclass[A4,11pt]{article}

\usepackage[T1]{fontenc}
\usepackage{authblk}
\usepackage[numbers,sort&compress]{natbib}

\usepackage{times}
\usepackage[left=1in, right=1in, top=1in, bottom=1in]{geometry}

\usepackage{amsmath,amssymb}

\usepackage{changepage}

\usepackage[utf8x]{inputenc}

\usepackage{textcomp,marvosym}

\usepackage[dvipsnames,table]{xcolor}

\usepackage{nameref,hyperref}

\usepackage{wrapfig}

\hypersetup{%
  colorlinks=true,
  linkcolor=blue,
  urlcolor=blue,
  citecolor=black,
  linkbordercolor=red,
}
\usepackage{microtype}
\DisableLigatures[f]{encoding = *, family = * }

\usepackage{array}

\usepackage{algorithm}
\usepackage{graphicx}
\usepackage[noend]{algpseudocode}
\usepackage[normalem]{ulem}
\usepackage{xspace}

\newcolumntype{+}{!{\vrule width 2pt}}

\newlength\savedwidth

\graphicspath{{../}}

\usepackage[aboveskip=1pt,labelfont=bf,labelsep=period,justification=raggedright,singlelinecheck=off]{caption}

\usepackage[dvipsnames,table]{xcolor}
\usepackage{ifthen}
\newboolean{draft}
\setboolean{draft}{false}
\ifthenelse{\boolean{draft}}{%
  \newcounter{comments}
  \newcommand{\anna}[1]{\addtocounter{comments}{1}{\color{blue}[AR \thecomments: #1]}}
    \newcommand{\larry}[1]{\addtocounter{comments}{1}{\color{cyan}[LZ \thecomments: #1]}}
      \newcommand{\jabriel}[1]{\addtocounter{comments}{1}{\color{BrickRed}[JZ \thecomments: #1]}}
        \newcommand{\gabe}[1]{\addtocounter{comments}{1}{\color{Emerald}[GP \thecomments: #1]}}
  \newcommand{\tobias}[1]{\addtocounter{comments}{1}{\color{RubineRed}[TR \thecomments: #1]}}
    \newcommand{\pramesh}[1]{\addtocounter{comments}{1}{\color{Plum}[PS \thecomments: #1]}}
    \newcommand{\new}[1]{{\color{red}#1}}
}{
\newcommand{\anna}[1]{}
\newcommand{\larry}[1]{}
\newcommand{\jabriel}[1]{}
\newcommand{\gabe}[1]{}
\newcommand{\tobias}[1]{}
\newcommand{\pramesh}[1]{}
\newcommand{\new}[1]{#1}
}

\usepackage{xspace}
\newcommand{\graphery}{\textsc{Graphery}\xspace}

\begin{document}

\title{\graphery: interactive tutorials for biological network algorithms}

\author[1,2]{Heyuan Zeng\thanks{zengl@reed.edu}}
\author[3]{Jinbiao Zhang}
\author[2]{Gabriel A. Preising}
\author[2]{Tobias Rubel}
\author[2]{Pramesh Singh}
\author[2]{Anna Ritz\thanks{aritz@reed.edu}}

\affil[1]{Computer Science Department, Reed College, Portland, OR, USA}
\affil[2]{Biology Department, Reed College, Portland, OR, USA}
\affil[3]{Information and Communication Technology Department, Xiamen University Malaysia, Selangor Darul Ehsan, Malaysia}
\date{}

\maketitle

\begin{abstract}
	Networks have been an excellent framework for modeling complex biological information, but the methodological details of network-based tools are often described for a technical audience. We have developed \graphery, an interactive tutorial webserver that illustrates foundational graph concepts frequently used in network-based methods. Each tutorial describes a graph concept along with executable Python code that can be interactively run on a graph. Users navigate each tutorial using their choice of real-world biological networks that highlight the diverse applications of network algorithms. \graphery also allows users to modify the code within each tutorial or write new programs, which all can be executed without requiring an account. \graphery accepts ideas for new tutorials and datasets that will be shaped by both computational and biological researchers, growing into a community-contributed learning platform. \graphery is available at \href{https://graphery.reedcompbio.org/}{https://graphery.reedcompbio.org/}.
\end{abstract}

\section{Introduction}

Computational biologists have long used networks, or \textit{graphs}, to model complex relationships that range in scale from molecular interactions to population and community dynamics. The popularity of graphs for bioinformatics and computational biology is marked by a wealth of review articles dedicated to the topic.  For example, graphs have been extensively used to model molecular interactions such as protein interactions, metabolic signaling, and gene regulatory relationships~\cite{aittokallio2006graph,cowen2017network,mcgillivray2018network,mitra2013integrative}. Network algorithms have also been developed to study the molecular systems biology of complex diseases, including identifying disease genes and pathways and predicting potential drug targets~\cite{cho2012network,creixell2015pathway}. Beyond molecular systems, graphs have been used to model neuron connections in the brain~\cite{bassett2018nature}, social dynamics in animals~\cite{sah2017unraveling,brugere2018network} and energy flux in populations (e.g., food webs)~\cite{girvan2002community,dunne2002food}.

Alongside this development of network-based methods, the amount of available 
network data (e.g., constructed from experiments or by carefully curating existing literature) has exploded.  Graph databases and repositories have been developed to store, query, and visualize biological networks~\cite{have2013graph,struck2020exploring,fabregat2018reactome,slenter2018wikipathways,pratt2015ndex,sah2019multi,mering2003string}, and biological network visualization is a subfield in its own right~\cite{gehlenborg2010visualization,hu2007towards}.  Web-based network visualization has enabled a lightweight and interactive means for users to explore graphs without downloading a stand-alone application~\cite{franz2016cytoscape,lopes2010cytoscape,bharadwaj2017graphspace}. Many graph algorithms that were designed for biological networks offer their own web-based tools, for example GeneMANIA~\cite{franz2018genemania} and SteinerNet~\cite{tuncbag2012steinernet}, or have plugins for existing platforms such as Cytoscape~\cite{lotia2013cytoscape}.  

Recent growth in biological network data has inspired novel graph algorithms that, for example, uncover important components of or identify higher-order organization within complex networks. 
Many of these methods are extensions of classic and well-studied graph algorithms, such as clustering and community detection, random walks and belief propagation, and finding paths and trees.  With the wealth of online datasets, web-based tools, and network graph visualization platforms, researchers can run these graph algorithms on their own data.  However, when it comes to interpreting the outputs of these algorithms, biological researchers often face a major obstacle: 
\textit{How can you interpret the outputs of an algorithm if you do not know how the algorithm works?}  

\enlargethispage{-65.1pt}

There has been growing recognition that it is important for biological researchers to understand the concepts that underpin current computational methodologies~\cite{carey2018ten,wilensky2014fostering,mulder2018development}. While review articles offer a broad view of an application or methodology, they tend to cite original work that may be dense for a non-computational audience.  Primers and workshops have been developed to give biologists a deeper understanding of mathematical concepts, including machine learning~\cite{mu2019ml4bio}, Bayesian network modeling~\cite{needham2007primer}, and deep learning~\cite{webb2018deep}.   There are also a few network-based primers and textbook chapters dedicated to giving biologists more insight into graph statistics and algorithms~\cite{dong2015reverse,pevzner2011bioinformatics,klipp2016systems,junker2011analysis}. While these are rich with examples and principles of graph algorithms, they do not tend to be interactive or designed for researchers who want a high-level understanding of graphs.

We present \graphery, a web-based platform that offers graph tutorials, example code, and interactive graphs, all in one place.  \graphery is aimed at helping biological researchers understand the core concepts upon which many state-of-the-art graph algorithms are built. 
The central idea behind \graphery is that users can read tutorials and run the associated Python code on real-world biological networks that may be relevant to their field (Supplementary Section~\ref{sec:overview}).
\graphery's power is in the user's ability to select a specific graph to use when working through the tutorials and run that tutorial's Python code on the graph using a step-through debugger-style interface.  



\section{\new{Audience}}

The majority of \graphery users will be \textit{visitors} who navigate the tutorials.  Visitors can interact with graphs, read through the tutorials, run code, and even edit the code without logging in.  \new{We expect most visitors to be trainees and researchers from the biological sciences that commonly use bioinformatic tools to analyze their data.  \graphery tutorials provide written descriptions of the fundamental concepts behind common graph algorithms and enable visitors to interact with real-world networks.  There is also a click-through interface that steps through Python code line-by-line, highlighting important variables.}  

\new{We hope that visitors will gain an intuition about the general control flow of a piece of code by stepping through it line-by-line. A smaller number of visitors may be more familiar with programming, and \graphery provides features for modifying or writing code alongside the existing tutorials and networks. However, teaching Python programming is outside the scope of our webserver.}

\new{In this paper,} we describe \graphery from a visitor perspective.  \new{We first present tutorials and graphs as a visitor would interact with them. We then provide details about other user roles (such as authors, translators, and administrators), \graphery's implementation, and the potential of \graphery for education and training. } 

\begin{figure}[h]
\includegraphics[width=\linewidth]{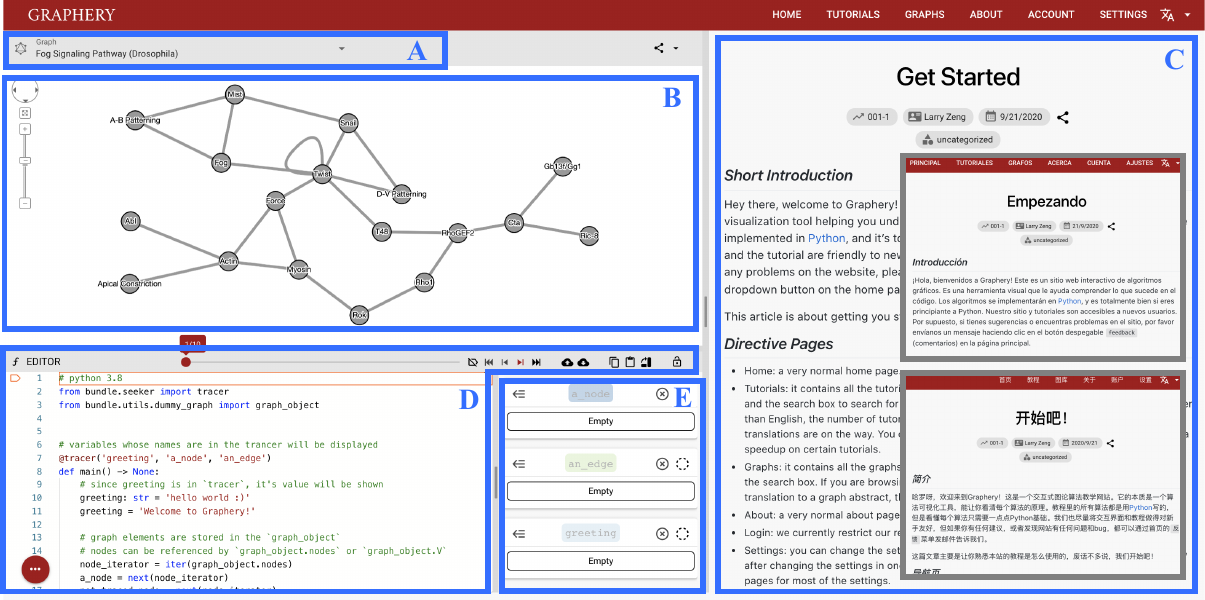}
\caption{Tutorial view. This page includes (A) a dropdown to visualize different graphs, (B) an interactive graph visualization, (C) tutorial content (which can be switched to Spanish or Chinese translations), (D) a debugger-style editor, and (E) a list of traced variables.}
\label{fig:tutorial-page}
\end{figure}

\section{GRAPHERY TUTORIALS}
\label{sec:tutorials}

Tutorials are the heart of \graphery, as they are the view where all features come together (Figure~\ref{fig:tutorial-page}).  
The right-hand pane of the tutorial view contains the \textit{tutorial's content}, including text, images, and hyperlinks (Figure~\ref{fig:tutorial-page}C).  Tutorials are ordered in increasing complexity, beginning with tutorials that orient users, moving on to simple definitions and characteristics of graphs, and then delving into the main concepts behind graph algorithms (Table~\ref{table:tutorials}).  We intentionally made these tutorials short and brief, with the goal that they build upon each other.  In some cases, later tutorials use code that is described in earlier tutorials.  
Many of our tutorials are written in English, Chinese, and Spanish (Table~\ref{table:tutorials} and inset of Figure~\ref{fig:tutorial-page}C), with a goal to add more languages that reach a broader audience of network biology researchers.

\begin{table}
\centering
\small
\begin{tabular}{|cll|ccc|} \hline
\textbf{Number} & \textbf{Name} & \textbf{Description} &\!\!\textbf{EN-US}\!\! &\! \!\textbf{ES}\!\! & \!\!\textbf{ZH-CN}\! \! \\ \hline\hline
001-1 & Get Started & Website structure and how to navigate tutorials & $\checkmark$ &  $\checkmark$&  $\checkmark$\\
\new{001-2} & \new{Graph Primer} & \new{Introduction to graphs in Graphery} & \new{$\checkmark$} & \new{$\checkmark$}&\new{$\checkmark$} \\
\new{001-3} & \new{Python Primer} & \new{Introduction to Python programming} & \new{$\checkmark$} &\new{$\checkmark$} & \new{$\checkmark$}\\
001-\new{4} & Programming in Graphery & Overview of the advanced program editor features & $\checkmark$ &  $\checkmark$& $\checkmark$ \\\hline
101-1 & Counting Elements & Introduction to graphs, nodes, and edges & $\checkmark$ &  $\checkmark$&  $\checkmark$\\
\new{101-2} & \new{Depth First Search} & \new{Introduction to depth first search traversal} & \new{$\checkmark$} & & \\ \hline
102-1 & Degree Distribution & \new{Degree distribution and network global structure} & $\checkmark$ & $\checkmark$ & $\checkmark$\\
102-2 & Degree Distribution 2 & Calculating the degree distribution, faster& $\checkmark$ &  $\checkmark$  & $\checkmark$\\\hline
103-1 & Shortest Paths & Computing the shortest path between two nodes& $\checkmark$ &  $\checkmark$ & $\checkmark$\\
103-2 & Shortest Paths 2 & Calculating the average shortest path length \new{of a graph}  & $\checkmark$ &  $\checkmark$ &$\checkmark$ \\
\new{103-3} & \new{Betweenness Centrality} & \new{Calculating the betweenness centrality for all nodes}  & \new{$\checkmark$} &  & \\\hline
104-1 & Trees \& Acyclic Graphs & Introduction to trees \new{and checking for graph acyclicity} & $\checkmark$ & $\checkmark$  &$\checkmark$ \\
104-2 & Spanning Trees & Spanning trees on unweighted graphs & $\checkmark$& &$\checkmark$\\ \hline
105-1 & Clustering Coefficient & Calculating a network's global clustering coefficient &$\checkmark$ & &\\ 
105-2 & Clustering Coefficient 2 & Calculating the local clustering coefficient of nodes  &$\checkmark$ & &\\  \hline
201-1 & Modularity & Community structure and network modularity& $\checkmark$ & &\\
201-2 & Modularity 2 & Greedy community detection by maximizing modularity& $\checkmark$& &\\  \hline
202-1 & Random Walks & Introduction to random walks &$\checkmark$ & &\\ 
202-2 & Random Walks 2 & Random walks with restarts &$\checkmark$ & &\\ \hline
\new{203-1} & \new{Label Propagation} & \new{Introduction to label propagation} & \new{$\checkmark$} & & \\\hline\hline
\end{tabular}
\caption{Tutorials available in \graphery, along with their available Spanish and Chinese translations (last updated \new{April 18, 2021}).}
\label{table:tutorials}
\end{table}

\graphery currently supports undirected, unweighted graphs.
The tutorials cover a number of fundamental concepts about undirected graphs, including graph definitions (such as nodes, edges, node degree, paths, and trees),  graph statistics and properties (such as degree distribution, clustering coefficient, and acyclicity), simple graph algorithms (such as shortest paths and spanning trees), and more advanced graph concepts (such as random walks and community detection).  These tutorials serve as a foundation for describing more complicated concepts, such as label propagation (e.g. used in GeneMania~\cite{franz2018genemania}), Steiner trees (e.g. used in SteinerNet~\cite{tuncbag2012steinernet}), and $k$-shortest paths (e.g., used in PathLinker~\cite{ritz2016pathways}). We are actively developing new \graphery tutorials, and we welcome suggestions and contributions.

\graphery contains an interactive \textit{graph visualization panel} (Figure~\ref{fig:tutorial-page}B) that is built upon Cytoscape.js~\cite{franz2016cytoscape}.  Visitors can manually rearrange nodes in the graph and use the pan and zoom features.  The drop-down in Figure~\ref{fig:tutorial-page}A allows a visitor to select among a list of available biological networks for this tutorial; the networks themselves are described in more detail in the next section.

\begin{wrapfigure}{r}{.55\linewidth}
\centering
\includegraphics[width=\linewidth]{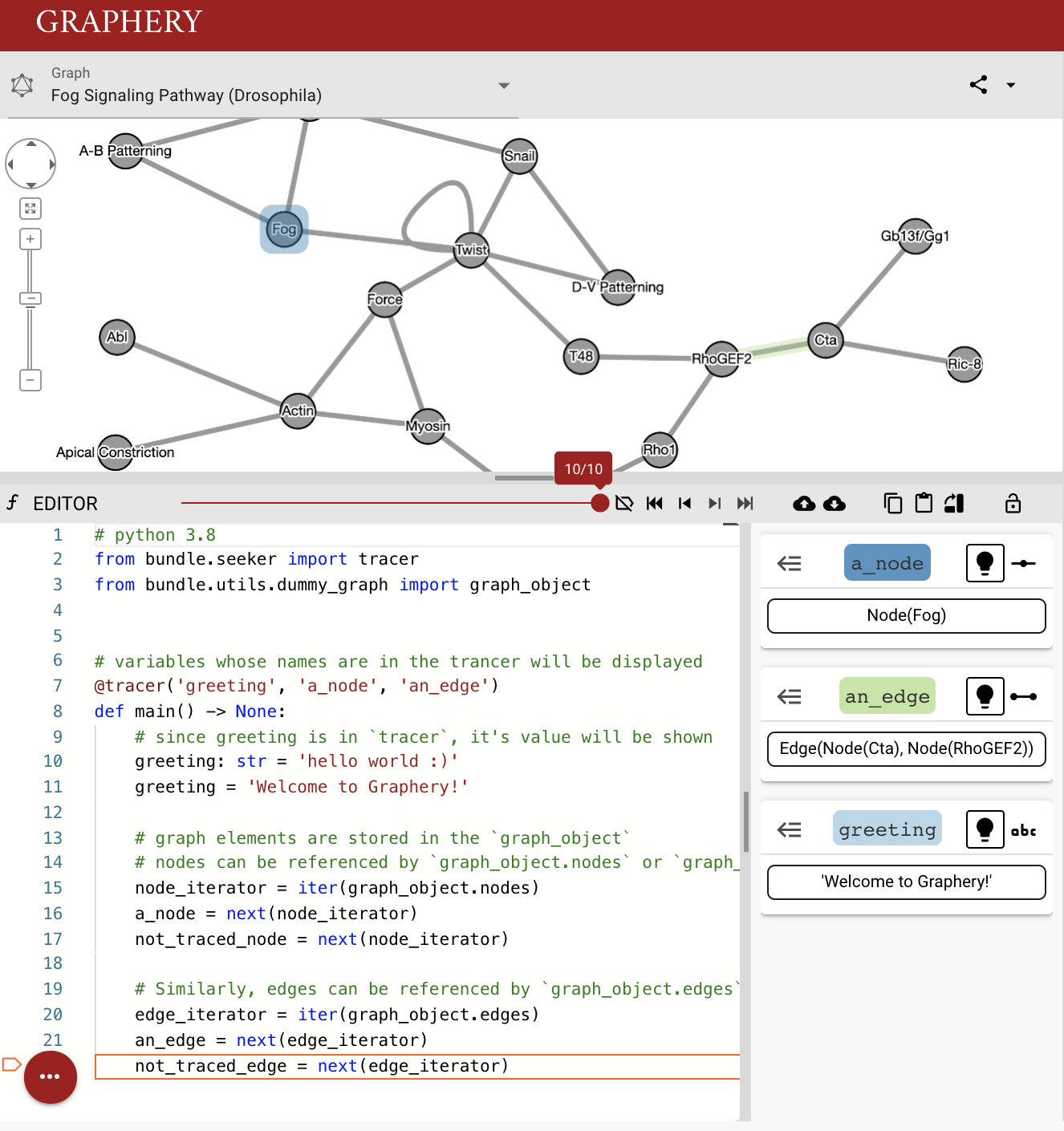}
\caption{Example of the code after execution (Panels D and E from Figure~\ref{fig:tutorial-page}).  The list of traced variables is now populated and graph variables are highlighted in the interactive panel.}
\label{fig:code-stepthrough}
\end{wrapfigure}

Visitors can click through each tutorial's Python code in the \textit{code editor}, which employs a debugger-style format (Figure~\ref{fig:tutorial-page}D and Supplementary Figure~\ref{fig:pySnooper})~\cite{rachum2019pysnooper}.  The tool bar allows visitors to slide the step-counter or take one- or five-step jumps with the next/reverse buttons.  Importantly, the code contains traced variables, which appear in a variable list (Figure~\ref{fig:tutorial-page}E) and are also highlighted in the graph itself.  Figure~\ref{fig:code-stepthrough} shows the result after running the code in the \textit{Get Started} tutorial, which selects a random node and a random edge in the graph. 

\new{Some visitors may be interested in learning more about how the Python code carries out the specified tasks.}  The code editor offers additional features for visitors to gain familiarity with programming in the context of the displayed tutorial.  
After unlocking the editor (by clicking the lock button on the toolbar or changing \graphery's settings), visitors can change the variables being traced, change hard-coded parameters to observe their effects, or write completely new functions within the editor.  Whenever the code is changed, visitors just need to execute the code \new{with a one-button click} and the new modifications will be applied. \new{\graphery provides remote execution capabilities (the cloud button with an up arrow), or local execution capabilities on the user's machine if a back-end server is running (the cloud button with a down arrow).} Through these features, \graphery visitors with different programming experience levels can learn from each tutorial. 

\section{Graphery Graphs}
\label{sec:graphs}

If tutorials are the heart of \graphery, biological networks are the cardiovascular system.  The tutorials would not be useful on their own without the ability for visitors to walk through the code using graphs. Further, these graphs are not only examples, but come from real-world datasets and applications.  Each graph contains a description of the underlying dataset (Supplementary Figures~\ref{fig:overview} and~\ref{fig:pySnooper}).  
The drop-down allows users to specify other graphs on which the Python code can be executed (Figure~\ref{fig:tutorial-page}A).  
\graphery also offers a playground feature where users can open a biological network with a code editor and interact with the graph independent of any specific tutorial (Supplementary Figure~\ref{fig:playground}). 


The biological networks in \graphery span different scales of biology, from molecular interactions to population dynamics (Table~\ref{table:networks}).  Further, they also capture networks relevant to different diseases (such as tuberculosis spread in a badger population, tumor cell evolution, and dysregulated protein networks).  There is a current focus on protein-protein interactions and signaling pathways due to the authors' research areas, and we are continuing to expand the selection of biological networks to other domains and network topologies. Just like tutorials, we are actively posting new \graphery graphs, and we welcome both suggestions and contributions.

\begin{table}[b]
\centering
\small
\begin{tabular}{|llc|ccc|} \hline
\textbf{Name} & \textbf{Network Type} & \textbf{Refs} & \!\!\textbf{EN-US}\!\! & \!\!\textbf{ES}\!\! & \!\!\textbf{ZH-CN}\!\!  \\ \hline \hline
Badger Social Network & disease\new{; social network} & \cite{weber2013badger} & $\checkmark$ &$\checkmark$ & \\ 
Colorectal Cancer Evolution Tree & disease\new{; evolution} &  \cite{dang2020clonal} & $\checkmark$ & $\checkmark$& \\
Competition Graph of Yellowstone Food Web  & ecology\new{; food web} & \cite{koirala2018food}  & $\checkmark$ & $\checkmark$& \\ 
Fog Signaling Pathway (\textit{fly})  & protein interactions\new{; pathway} & \cite{manning2014fog,peters2018cell} & $\checkmark$ & $\checkmark$& $\checkmark$\\ 
Food Web of Intertidal Species in WA & ecology\new{; food web} & \cite{hui2012food} & $\checkmark$ &$\checkmark$ & $\checkmark$\\
Interleukin-9 Signaling Pathway (\textit{human}) & protein interactions\new{; pathway} & \cite{kandasamy2010netpath}& $\checkmark$ & $\checkmark$& \\ 
Pan-Cancer Network & disease\new{; protein interactions} & \cite{leiserson2015pan} &$\checkmark$ &$\checkmark$ & \\ 
Protein Transport Complex (\textit{fly}) & protein interactions & \cite{guruharsha2011protein}& $\checkmark$ & &  \\
Tutorial Network & \multicolumn{2}{l|}{\textit{base network for tutorials}} & $\checkmark$ &$\checkmark$ &$\checkmark$\\ \hline
\end{tabular}
\caption{Biological networks available as \graphery graphs (last updated \new{April 18, 2021}).}
\label{table:networks}
\end{table}

\section{Other Roles in Graphery}
\label{sec:roles}

While visitors are the most common type of users, \graphery provides other roles that support the development and maintenance of the webserver.  \textit{Authors} are users who contribute new tutorials and/or graphs.  Authors have control over which graphs are utilized in their tutorial, tutorial/network graph categories, and whether their content is published (viewable by visitors).  \textit{Translators} are users who write translations of the existing tutorials.  Translators have flexibility in their translated content, for example by including references to additional or alternate sources that are more readily accessible in some countries.
\textit{Administrators} have control over all content published on \graphery, including those published by other authors and translators.  These user roles require an account, which is maintained by an administrator.   

Authors, translators, and administrators add content to \graphery through a user-friendly control panel (Supplementary Section~\ref{sec:roles} \new{ and Supplementary Figure~\ref{fig:control-panel}}).  The control panel supports image upload, Markdown-style editing with a live-update preview mode, and summaries of the posted tutorials and graphs.  Authors and translators receive attribution on the content they contribute, as seen by the author tag in Figure~\ref{fig:tutorial-page}.  

Any user, including contributors such as authors/translators or visitors who interact with tutorials, can run Python code locally instead of in the cloud. Running the code locally gives users more control over the execution process, for example working with local files and using 
external Python modules.  Supplementary Section~\ref{sec:implementation} \new{ and Table~\ref{tab:software}} provides more implementation details and instructions for local execution. \new{The code is available on GitHub at \href{https://github.com/Reed-CompBio/Graphery}{https://github.com/Reed-CompBio/Graphery}.}

\section{\new{Graphery for Education and Training}}
\label{sec:education}

\new{\graphery has potential to be adapted for computational biology education and training at multiple levels, from high school students to bioinformaticians new to systems biology.  In its current state, we believe that the majority of \graphery visitors will be (a) trainees in the biological sciences who may commonly use tools with graph algorithms (b) undergraduate science majors without substantial programming experience. With these groups of trainees in mind, we considered core competencies described by two overlapping but distinct societies: the Curriculum Task Force of the International Society for Computational Biology (ISCB) Education Committee~\cite{mulder2018development} and The Network for Integrating Bioinformatics into Life Sciences Education (NIBLSE)~\cite{wilson2018bioinformatics}.  ISCB's competencies were focus on bioinformatics training, whereas NIBLSE focus on computational competencies designed for trainees in the life sciences. We mapped common \graphery tasks (reading tutorials, interacting with graphs and code, modifying code, and writing code) with core competencies laid out by ISMB and NIBLSE (Supplementary Figure~\ref{fig:competencies}).  All visitors will be able to evaluate graph algorithms in the context of systems biology and use bioinformatics tools to explore biological interactions and networks.  As visitors engage more with the tutorials by learning about the code, they will begin to understand the algorithms and try adapting them with minor changes to the code. More experienced visitors who write new programs will learn about computing requirements needed to solve a problem. We note that there is no assessment in place for \graphery tutorials, so carefully designed user studies are needed to fully determine \graphery's potential in education and training.}

\section{DISCUSSION}
\label{sec:discussion}

\graphery provides an interactive environment to learn about concepts in graph algorithms that form the foundation of many state-of-the-art methods. The combination of tutorials, code, and real-world graphs encourages learning in a more engaging way than the tutorials alone.  We believe that \graphery has the potential to become a mainstay learning tool for biological researchers at any career stage, from students to principal investigators. 

We have come across some limitations in implementing our first tutorials and graphs.  First, \graphery visualizes undirected, unweighted graphs, which we acknowledge is only a subset of the type of graphs used in biological applications.  We plan to extend \graphery to support directed graphs, weighted graphs, and multigraphs (all of which Cytoscape.js supports). Second, we have found that even classic graph algorithms require a fair amount of background knowledge. While the tutorial content can describe this knowledge, the associated code must not incur an unreasonable number of steps for the code editor.  For example, while graph clustering is intuitive to describe at a high level, the code to cluster graphs (and trace the relevant variables) may end up taking hundreds of thousands of steps, even for the smallest graphs. Our approach to this challenge is to ensure that the tutorials describe small, modular components of algorithms that can be combined for later tutorials.  We also plan to implement additional useful step jumping tools to the execution set to help navigate code with many steps.  Finally, we believe that the real-world biological networks in \graphery are a fantastic way for biologists to better understand graph algorithms, but finding small and relevant graphs that describe real biological systems has been more difficult than we initially expected.  Some of our networks have been suggested by biological experts, which has proved to be a successful way to add graphs to \graphery.

In addition to adding new graphs and tutorials, our team has plans for larger updates to help users better understand the provided code.  The current graph API was originally developed for \graphery tutorials, and users who wish to modify or write code must have a basic understanding of object oriented programming and some knowledge of the underlying graph objects. While this information is described in the \textit{Programming in Graphery} tutorial and on our documentation webpage (Supplementary Section~\ref{sec:implementation}), we plan to swap our custom API with a graph package such as \texttt{networkx}~\cite{hagberg2008exploring}. This will help users already familiar with \texttt{networkx} to be able to modify code easily, and there will be a large amount of existing resources for programming support.  While \graphery is not explicitly designed to teach users how to program in Python, this modification will help users understand the programs provided in the tutorials.

Lastly, \graphery will become a better webserver with more involvement from the network biology community.  In addition to suggesting new tutorials and graphs, future versions will provide citations for current papers that use the concepts in each tutorial.  We welcome anyone who wishes to contribute new content for the website; they will be acknowledged as an author on the networks and tutorials. \graphery has potential to reach a broad audience through the translated pages, and we are actively seeking more translators to help us with this task (especially for languages that are not yet in place). With these goals in mind, \graphery will help bridge the computational and biological worlds of systems biology. 

\section{Acknowledgements}

This work was supported by the National Science Foundation (BIO-ABI \#1750981 to AR \new{and BIO-MCB \#1716964 to AR as co-PI}). GAP was supported by a National Institutes of Health supplement PA-20-222 to (NIH-NIGMS \#R15-DK116224-01 to SCP Renn).

\bibliographystyle{natbib}
\bibliography{zeng_graphery}

\appendix
\renewcommand\thefigure{\thesection\arabic{figure}}  
\renewcommand\thetable{\thesection\arabic{table}}  
\setcounter{figure}{0}  
\setcounter{table}{0}  

\section{Supplementary Information: Graphery Overview}
\label{sec:overview}
\begin{figure}[h]
\centering
\includegraphics[width=.8\linewidth]{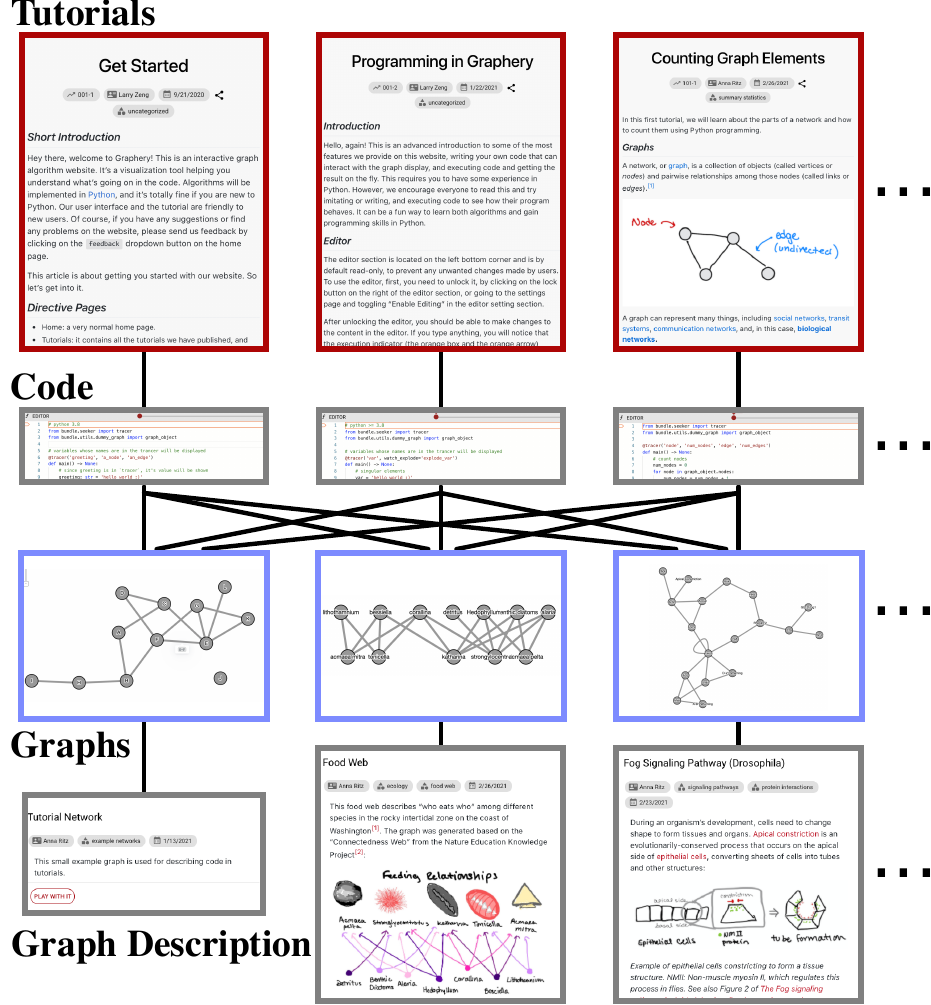}
\caption{Overview of \graphery.  Tutorials and code (top) are connected to graphs and descriptions (bottom), where most graphs can be used for most tutorials. Refer to the Section~\ref{sec:overview} for more details.  }
\label{fig:overview}
\end{figure}

Figure~\ref{fig:overview} provides a comprehensive view of \graphery organization. There are two main pieces: tutorials and graphs.  Tutorials are Markdown-style documents, which may contain hyperlinks, figures, and references.  Each tutorial has an associated piece of Python code, which the user can run using a step-through, debugger-style interface.  Graphs are real-world networks that are visualized using Cytoscape.js, and contain information about that graph (typically some details about the type of graph and relevant references). Graph descriptions are also Markdown-style documents.  Tutorial authors determine which graphs are appropriate for the Python code; typically, all graphs in the database are linked to each tutorial, represented by the multi-way connections in the middle of the figure.

\begin{figure}[h]
\includegraphics[width=\linewidth]{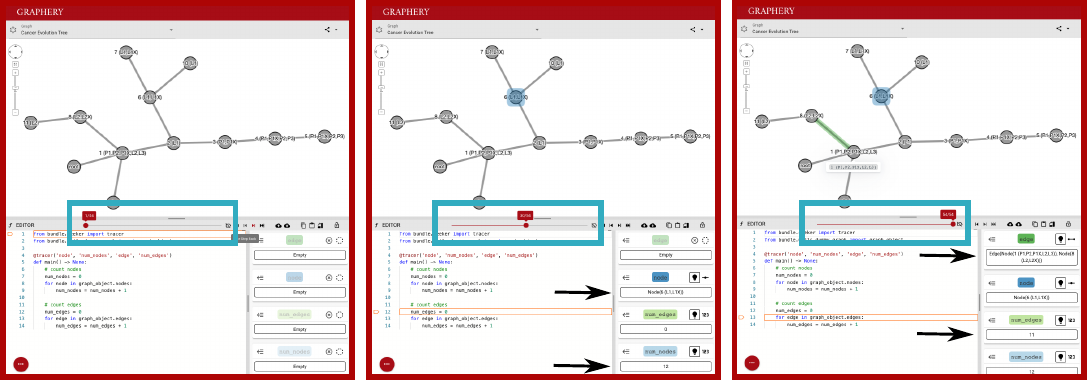}
\caption{Example of the debugger-like code interface for the \textit{Counting Nodes and Edges} tutorial on the Cancer Tumor Evolution network. (Left) The code before execution (denoted by the slider in the cyan box). (Middle) The code partway through execution after counting all nodes. The highlighted node is assigned to the `node` variable. Arrows indicate updated traced variables. (Right) The code after execution, after counting all nodes and edges. }
\label{fig:pySnooper}
\end{figure}

\begin{figure}[h]
\centering
\includegraphics[width=\linewidth]{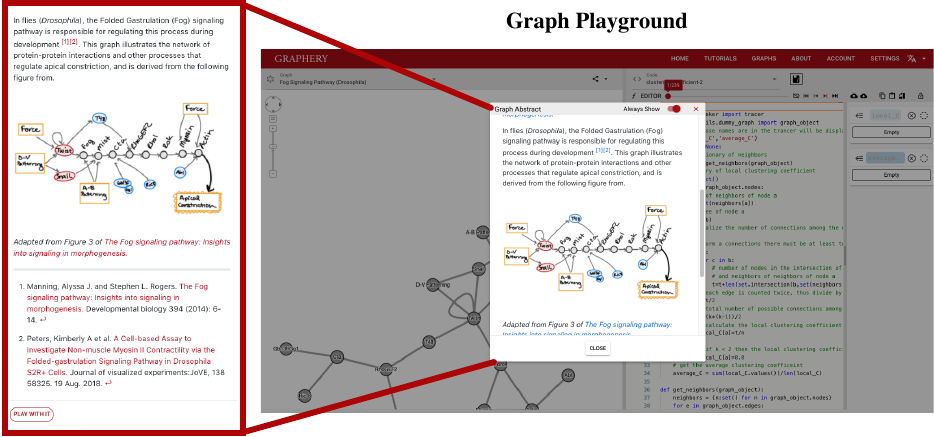}
\caption{Overview of the Graph Playground.  Each network has an associated graph description, or \textit{abstract}, that contains a brief description, images, and clickable references for the network.  Clicking the ``Play with It'' button in the Graphs tab brings the user to the graph playground, which contains the graph and a drop-down with all existing code available for that graph.}
\label{fig:playground}
\end{figure}

\clearpage
\section{Supplementary Information: Authors, Translators, \& Administrators}
\label{sec:roles}

While visitors are the most common type of users, \graphery offers other roles that support the development and maintenance of the webserver.  \textit{Authors} are users who contribute new tutorials and/or graphs.  Tutorial authors submit Markdown-style tutorial content and Python code that links to the tutorial.  Authors of graphs submit JSON-formatted graphs and a Markdown-style description of the graph.  Authors have control over which graphs are displayed for their tutorial, tutorial/network graph categories, and whether their content is published (viewable by visitors).  \textit{Translators} are users who write translations of the existing tutorials, again in Markdown-style content.  Translators have flexibility in their translated content, for example by including references to additional or alternate sources that are more readily accessible in some countries.

\begin{figure}[th]
\includegraphics[width=\linewidth]{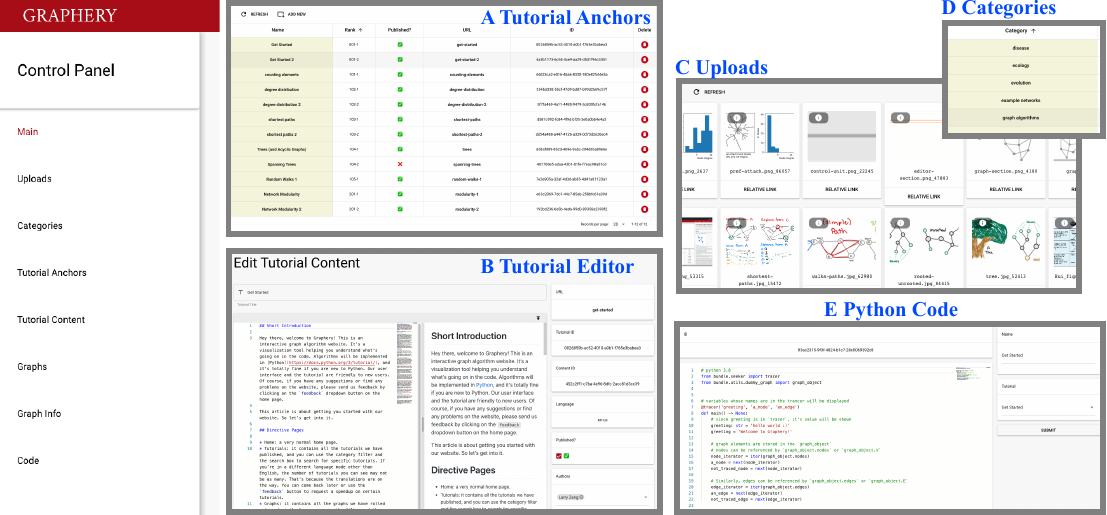}
\caption{Control panel for adding and modifying tutorial and graph content.  Panels A--E show the components used for developing a tutorial.  (A) Each tutorial requires an anchor, which specifies its unique hyperlink, its ranking, and whether it is published (viewable to visitors).  (B) The editor for a single tutorial includes a Markdown editor with an automatic preview.  (C,D) Contributors can upload images and create new categories for the tutorial. (E) Finally, code is linked to the tutorial.}
\label{fig:control-panel}
\end{figure}

Authors, translators, and administrators add content to \graphery through a user-friendly control panel. Within the control panel, users develop tutorials by first creating an anchor, then adding Markdown content, and finally linking Python code to the tutorial  (Figure~\ref{fig:control-panel}).  Translators can add Markdown content to a tutorial once it exists in English.  Users also contribute graphs through the control panel, first by uploading the graph itself as a JSON file and then adding a description for the graph in Markdown.  The control panel also indicates which tutorials and graphs have been published, and are viewable by visitors (for example, there is one tutorial in Figure~\ref{fig:control-panel}A that is not yet published).

We have provided documentation different user roles on \href{https://docs.graphery.reedcompbio.org}{https://docs.graphery.reedcompbio.org}.  For writers (e.g., authors and translators), we have a Getting Started page.  There is also a detailed User Manual for the data structures and objects available for programmers and the JSON file format specification for uploaded networks.

\clearpage
\section{Supplementary Information: Implementation}
\label{sec:implementation}
\graphery's front end is primarily written in Javascript using Quasar, which is built upon Vue.js. The tutorial page is powered by additional libraries such as Cytoscape.js~\cite{franz2016cytoscape} for network visualization, the Monaco Editor for code display, and Markdown It for Markdown text display.  The back end is written in Django and uses a PostgreSQL database for data storage, following a GraphQL API specification.
These libraries are provided in Table~\ref{tab:software} and others are listed on \graphery's \textit{About} page.

\graphery's code tracing features are based on a modified version of PySnooper~\cite{rachum2019pysnooper}.
After each line in the Python editor is executed, a custom trace function receives the stack frame of the current line. Our modified PySnooper records the changes made to traced variables and passes this information on to the front end to highlight the appropriate nodes and edges in the graph. These changes are also displayed in the list of traced variables (Figure~\ref{fig:pySnooper}).

\begin{table}
\centering
\begin{tabular}{|r|l|} \hline
\textbf{Library} & \textbf{URL} \\ \hline
Quasar & \href{https://quasar.dev/}{https://quasar.dev/} \\
Vue.js &  \href{https://vuejs.org/}{https://vuejs.org/}  \\
Cytoscape.js &  \href{https://js.cytoscape.org/}{https://js.cytoscape.org/} \\
Monaco Editor & \href{https://microsoft.github.io/monaco-editor/}{https://microsoft.github.io/monaco-editor/}  \\
Markdown It & \href{https://github.com/markdown-it/markdown-it}{https://github.com/markdown-it/markdown-it}  \\
Django & \href{https://www.djangoproject.com/}{https://www.djangoproject.com/} \\
GraphQL API & \href{https://graphql.org/}{https://graphql.org/}  \\
PySnooper & \href{https://github.com/cool-RR/PySnooper}{https://github.com/cool-RR/PySnooper}   \\\hline
\end{tabular}
\caption{Libraries and software used in \graphery's implementation.}
\label{tab:software}
\end{table}

Users may want to run their Python code locally instead of in the cloud. Running the code locally will give users more control over the execution process, for example working with local files and using external packages.  To locally start an execution server, you must download and run a simple Python script from the most recent release or the GitHub repository: 

\begin{center}\href{https://github.com/Reed-CompBio/Graphery/releases}{https://github.com/Reed-CompBio/Graphery/releases}.\end{center}

We will briefly describe how to run the local server from the most recent release (Python 3.7 or above is required to run the server).  Go the release page (\href{https://github.com/Reed-CompBio/Graphery/releases}{https://github.com/Reed-CompBio/Graphery/releases}) and download the latest \texttt{user\_server.zip} file. Extract everything in the zip file to a folder.  Move to the folder you extracted and execute \texttt{python user\_server.py}.  More details about the local Python execution can be found on the Documentation website: 

\begin{center}\href{https://docs.graphery.reedcompbio.org/user-manual/local-server/}{https://docs.graphery.reedcompbio.org/user-manual/local-server/}. \end{center}

\clearpage
\section{Supplementary Information: Core Competencies}
\label{sec:competencies}

\begin{figure}[h]
\centering
\includegraphics[width=.75\linewidth]{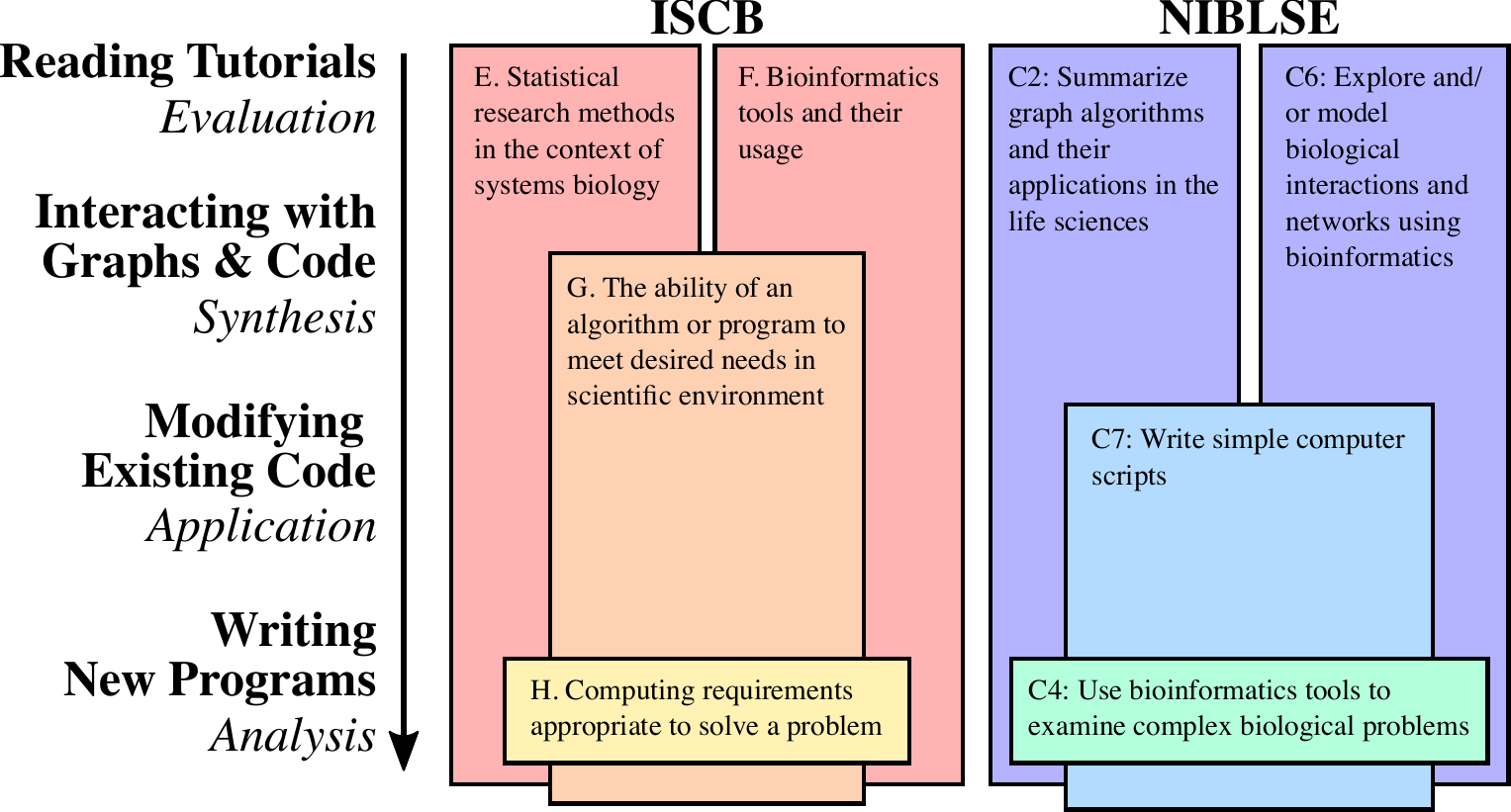}
\caption{Core competencies related to \graphery, according to the Curriculum Task Force of the International Society for Computational Biology (ISCB) Education Committee~\cite{mulder2018development} and The Network for Integrating Bioinformatics into Life Sciences Education (NIBLSE)~\cite{wilson2018bioinformatics}.  Graphery activities requiring increasing engagement with graphs and graph algorithms are shown on the left, with Bloom's Taxonomy terms in italics.  Relevant core competencies from ISCB are shown in red/yellow and from NIBLES are shown in blue/green. We anticipate that most Graphery visitors will read tutorials and interact with graphs and code. However, many Graphery visitors stop short of modifying existing code or writing new programs.}
\label{fig:competencies}
\end{figure}

\end{document}